\documentclass[runningheads]{llncs}
\usepackage{color}
\usepackage{graphicx}
\usepackage{amsmath}
\usepackage{amssymb}
\usepackage{booktabs}
\usepackage{multirow}
\usepackage{bm}
\usepackage[misc]{ifsym}
\usepackage[
colorlinks,
linkcolor=blue,
anchorcolor=blue,
filecolor=blue,
urlcolor=blue,
citecolor=blue]{hyperref}
\usepackage[capitalise]{cleveref}
\Crefname{section}{Sec.}{Secs.}
\usepackage[T1]{fontenc}
\usepackage[marginal]{footmisc}

\begin{document}
\title{4DRGS: 4D Radiative Gaussian Splatting for Efficient 3D Vessel Reconstruction from Sparse-View Dynamic DSA Images}
\titlerunning{4DRGS: 4D Radiative Gaussian Splatting
}
%
\author{
Zhentao Liu\inst{1*} \and
Ruyi Zha\inst{2*} \and
Huangxuan Zhao\inst{3} \and
Hongdong Li\inst{2} \and \\
Zhiming Cui \inst{1}(\Letter)
}

%
\authorrunning{
  Z. Liu et al.
}
%
\institute{
School of Biomedical Engineering \& State Key Laboratory of Advanced Medical Materials and Devices, ShanghaiTech University, Shanghai, China\\
\email{cuizhm@shanghaitech.edu.cn}\and
School of Computing, Australian National University, Canberra, Australia\and
School of Computer Science, Wuhan University, Wuhan, China
}

\maketitle              
\footnote{* Equal contribution.}
\begin{abstract}

Reconstructing 3D vessel structures from sparse-view dynamic digital subtraction angiography (DSA) images enables accurate medical assessment while reducing radiation exposure.
Existing methods often produce suboptimal results or require excessive computation time.
In this work, we propose 4D radiative Gaussian splatting (4DRGS) to achieve high-quality reconstruction efficiently.
In detail, we represent the vessels with 4D radiative Gaussian kernels.
Each kernel has time-invariant geometry parameters, including position, rotation, and scale, to model static vessel structures.
The time-dependent central attenuation of each kernel is predicted from a compact neural network to capture the temporal varying response of contrast agent flow.
We splat these Gaussian kernels to synthesize DSA images via X-ray rasterization and optimize the model with real captured ones.
The final 3D vessel volume is voxelized from the well-trained kernels.
Moreover, we introduce accumulated attenuation pruning and bounded scaling activation to improve reconstruction quality.
Extensive experiments on real-world patient data demonstrate that 4DRGS achieves impressive results in 5 minutes training, which is 32$\times$ faster than the state-of-the-art method.
This underscores the potential of 4DRGS for real-world clinics.

\end{abstract}
%

\section{Introduction}

\begin{figure*}[t]
\centering
    \includegraphics[width=1.0\textwidth]{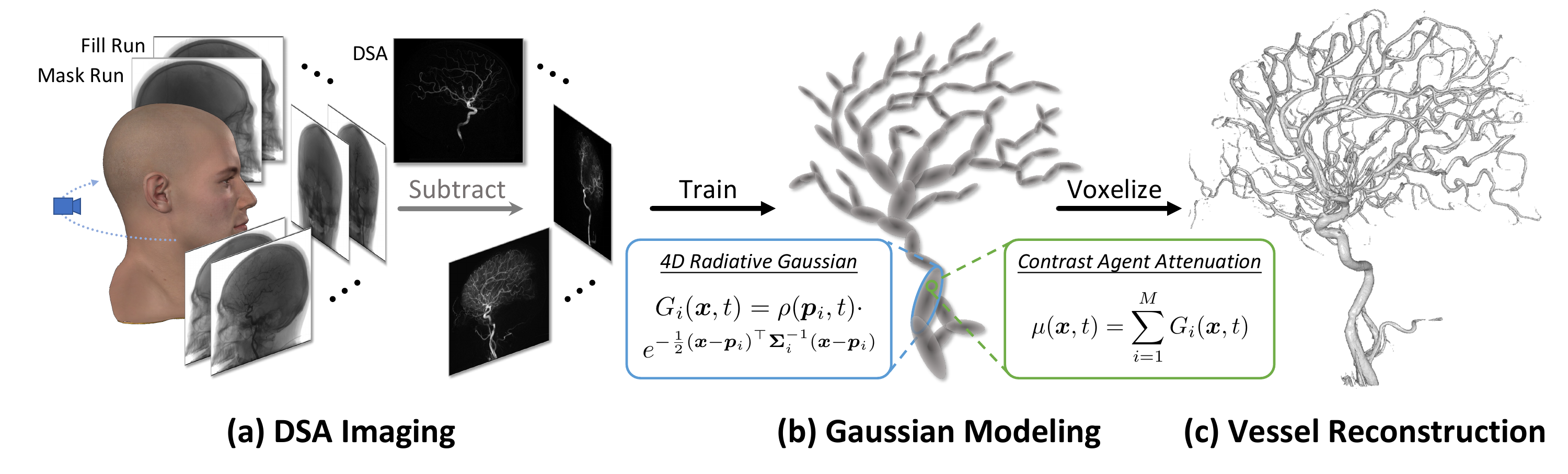}
    \caption{
    Overview of DSA imaging and vessel reconstruction. 
    (a) DSA images are generated by subtracting fill-run X-ray images from their mask-run counterparts.
    (b) We model vessels as a set of 4D radiative Gaussians.
    (c) The final 3D vessel volume is reconstructed via attenuation voxelization.
    }
    \label{fig:intro}
\end{figure*}

Digital subtraction angiography (DSA) is a widely recognized gold standard for diagnosing vascular diseases, such as arteriovenous malformation, arteriovenous fistula, and intracranial aneurysms~\cite{DSA_app_1,DSA_app_2,DSA_app_3,DSA_app_4}.
As shown in \cref{fig:intro}(a), DSA imaging involves two rotational cone-beam X-ray scans: the mask run, performed before the contrast agent injection, and the fill run, performed after the injection.
Subtracting X-ray images in the fill run from those in the mask run yields 2D DSA images, which highlight blood flow marked by the contrast agent while removing background tissues.
However, significant vessel overlap in DSA images hinders accurate anatomical assessment.
Therefore, reconstructing 3D vessel structures from DSA images is essential for clear visualization to support medical diagnosis.
Existing DSA systems typically capture hundreds of images for precise reconstruction~\cite{3DDSArecon} based on the Feldkamp-Davis-Kress (FDK) algorithm~\cite{FDK,fdk-3ddsarecon}, exposing patients and radiographers to significant radiation.
In this work, we aim to achieve high-quality reconstruction efficiently with dozens of images to reduce radiation exposure.

Sparse-view DSA reconstruction is a challenging task for two reasons.
First, as observed from \cref{fig:intro}(a), each DSA image captures a distinct blood state as the contrast agent gradually flows through the vessels.
Static computed tomography (CT) reconstruction algorithms~\cite{FDK,SART,ASDPOCS,Intratomo,NeAT,NAF,SNAF,sax_nerf,R2GS} often fail under such dynamic scenarios.
Second, sparse-view reconstruction is highly ill-posed.
Commonly used algorithms such as FDK~\cite{FDK,fdk-3ddsarecon} would produce severe artifacts when measurements are insufficient.
Although recent neural radiance fields (NeRF)~\cite{NeRF}-based method VPAL~\cite{VPAL} addresses these challenges well, it takes hours to process a single case and is thus impractical for real-world usage.

Our work is inspired by 3D Gaussian splatting (3DGS)~\cite{3DGS}, which represents scenes using explicit kernels and employs rasterization for RGB image rendering.
Compared with NeRF-based methods~\cite{VPAL} which model the entire space using neural networks, the kernel-based representation allows us to bypass large empty regions in backgrounds and focus on sparse vessel structures. 
Furthermore, differentiable rasterization offers faster rendering and training than volume rendering~\cite{VPAL}, making it well-suited for time-sensitive DSA reconstruction.
Existing 3DGS-based solutions in X-ray imaging address static CT reconstruction~\cite{ligsct,R2GS}, X-ray image synthesis~\cite{X-Gaussian,DDGS-CT}, and DSA image synthesis~\cite{TOGS}.
However, no prior work has applied Gaussian splatting to DSA reconstruction.

In this paper, we introduce 4D radiative Gaussian splatting (4DRGS), the first Gaussian splatting-based framework for efficient 3D vessel reconstruction from sparse-view dynamic DSA images.
A key observation is that vessels maintain static structures during scanning process, while their attenuation varies over time due to the contrast agent flow.
Therefore, we represent vessels as a set of 4D radiative Gaussian kernels (\cref{fig:intro}(b)), where each kernel acts as a local Gaussian-shaped time-varying attenuation distribution.
The static vessel structures are modeled with time-invariant geometry parameters, including position, rotation, and scale.
To mimic the temporal attenuation changes, we use a compact neural network to predict each kernel's central attenuation based on its position and the given timestamp.
We splat these Gaussian kernels to synthesize DSA images via X-ray rasterization~\cite{R2GS} and optimize them by minimizing the disparities with real captured images.
After training, 3D vessel volume is reconstructed via attenuation voxelization (\cref{fig:intro}(c))~\cite{R2GS}.
Two innovations are further introduced to enhance reconstruction quality: (1) accumulated attenuation pruning to remove non-vessel kernels and (2) bounded scaling activation to reduce needle artifacts. 
Experiments on real-world data demonstrate our method's effectiveness for both 3D vessel reconstruction and 2D DSA image synthesis.
Notably, 4DRGS achieves impressive results in 5 minutes and converges in 13 minutes, offering a speedup of 32$\times$ and 12$\times$ compared to the state-of-the-art (SOTA) method VPAL~\cite{VPAL}.

In summary, our contributions are threefold. 
First, we propose 4DRGS, the first Gaussian splatting-based framework for efficient 3D vessel reconstruction. 
Second, we develop 4D radiative Gaussian kernels for DSA imaging and introduce key innovations to improve reconstruction quality. 
Finally, our method achieves SOTA results within minutes, demonstrating its potential for real-world usage.

\section{Preliminary}

\subsubsection{DSA Imaging}
\label{sec:DSA imaging}

As depicted in \cref{fig:intro}(a), DSA imaging involves two scans: one before injecting the contrast agent (mask run) and one after (fill run). 
Both scans follow the same procedure, where a cone-beam X-ray machine rotates around the patient and captures 2D X-ray images at equal angular intervals. 
Denote X-ray images from the mask run as $\{\mathbf{I}_{j}^{m}\in \mathbb{R}^{H\times W}\}_{j=1}^{T}$ and their counterparts from the fill run as $\{\mathbf{I}_{j}^{f}\in \mathbb{R}^{H\times W}\}_{j=1}^{T}$, where $j$, $T$, and $H\times W$ are frame index, total number of frames, and image resolution, respectively. 
The DSA images $\{\mathbf{I}_{j}\in \mathbb{R}^{H\times W}\}_{j=1}^{T}$ can be generated through logarithmic subtraction: $\mathbf{I}_j = \ln(\mathbf{I}_j^{m}) - \ln(\mathbf{I}_j^{f})$. 
In this way, DSA images highlight dynamic blood flows while removing non-relevant tissues,
delivering useful insights for vascular disease diagnosis.

\subsubsection{Sparse-View DSA Reconstruction}
\label{sec:sparse_view_DSA_reconstruction}

We define the timestamp $t_j$ of the $j$-th DSA image as $t_j=\frac{j}{T}$, which indicates its capture order. 
The complete set of frame data is represented as $\{ \mathbf{I}_j, t_j \}_{j=1}^{T}$. 
The goal of sparse-view DSA reconstruction is to recover an attenuation volume representing vessel structures with a uniformly sampled subset $\{ \mathbf{I}_{j_k}, t_{j_k}  \}_{k=1}^{N}$, where $N<T$ and $j_k =\lfloor (k-1)\cdot\frac{T}{N} \rfloor+1$.

\subsubsection{Radiative Gaussian Splatting}

R$^2$-Gaussian~\cite{R2GS} is the first work to leverage 3DGS for static CT reconstruction. It represents the scanned scene using 3D radiative Gaussian kernels, each defined by the central attenuation, position, rotation, and scale. It develops differentiable X-ray rasterization and voxelization to produce X-ray images and attenuation volumes in real time. 
Building on R$^2$-Gaussian, we introduce 4DRGS designated for dynamic DSA imaging, enabling high-quality and efficient vessel reconstruction.

\section{4D Radiative Gaussian Splatting}

\begin{figure*}[t]
\centering
    \includegraphics[width=1.0\textwidth]{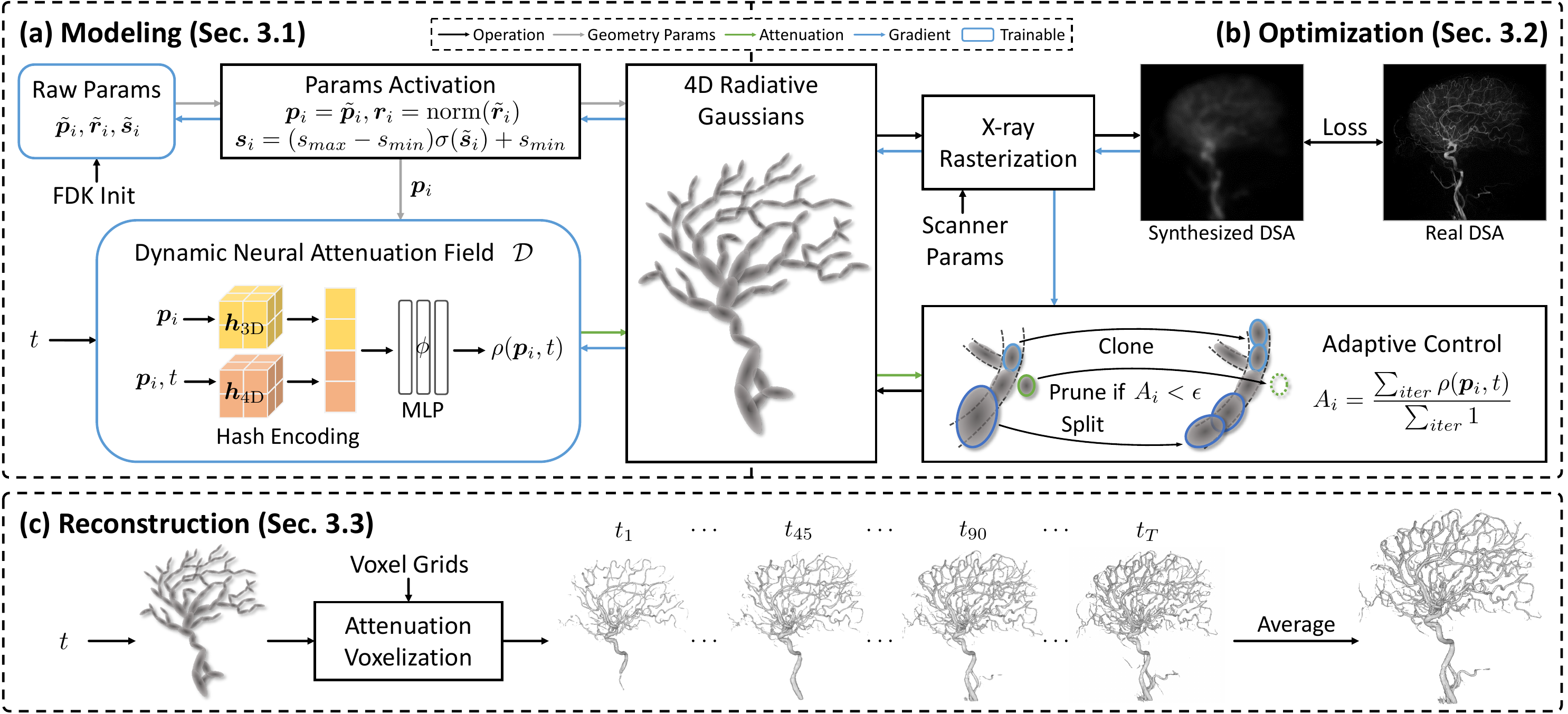}
    \caption{
    The overall pipeline of 4DRGS.
    We model vessels as a set of 4D radiative Gaussian kernels (\cref{sec:4d_Radiative_Gaussian_Modeling}) and optimize them with image losses (\cref{sec:model_optimization}). 3D vessel volume is reconstructed via attenuation voxelization (\cref{sec:vascular_reconstruction}).
    } 
    \label{fig:method}
\end{figure*}

The overall pipeline of 4DRGS is shown in \cref{fig:method}. 
First, we present 4D radiative Gaussian kernels as DSA scene representation in \cref{sec:4d_Radiative_Gaussian_Modeling}. 
Our 4D kernels extend the static 3D kernels in~\cite{R2GS} by modeling time-varying contrast agent flow and incorporating improved activation modules. 
Next, in \cref{sec:model_optimization}, we outline the training process and introduce accumulated attenuation pruning, which is designed for DSA imaging to improve reconstruction quality. 
Finally, we detail the process of extracting 3D vessels from the trained kernels in \cref{sec:vascular_reconstruction}.

\subsection{4D Radiative Gaussian Modeling}
\label{sec:4d_Radiative_Gaussian_Modeling}

\subsubsection{DSA Scene Representation}

As shown in~\cref{fig:intro}(b) and~\cref{fig:method}(a), we model vessels with a set of 4D radiative Gaussian kernels $\mathbb{G}=\{G_i\}_{i=1}^{M}$, where $M$ is the number of kernels. 
A key observation is that vessels maintain static structures during the scanning process, but their attenuation varies over time due to the flowing contrast agents.
Based on this, we define each kernel with two types of attributes including geometry parameters and central attenuation.
Time-invariant geometry parameters describe static vessel structures.
These parameters include position $\bm{p}_i\in\mathbb{R}^3$ for kernel location, rotation quaternion $\bm{r}_i\in\mathbb{R}^4$ for kernel orientation, and scale vector $\bm{s}_i\in\mathbb{R}_{+}^{3}$ for kernel size.
The rotation quaternion $\bm{r}_i$ and scale vector $\bm{s}_i$ can be further converted into rotation matrix $\mathbf{R}_i \in \mathrm{SO(3)}$ and diagonal scale matrix $\mathbf{S}_i \in \mathbb{R}^{3 \times 3}$, respectively, yielding the Gaussian covariance matrix $\mathbf{\Sigma}_i = \mathbf{R}_i \mathbf{S}_i \mathbf{S}_i^{\top} \mathbf{R}_i^{\top}\in\mathbb{R}^{3\times3}$~\cite{3DGS}. 
The central attenuation $\rho(\bm{p}_i, t)\in\mathbb{R}_{\geq 0}$ is time-dependent to capture attenuation changes caused by the contrast agent flow.
Finally, the Gaussian-shaped kernel response $G_i(\bm{x}, t) \in \mathbb{R}_{\geq 0}$ for any spatial point $\bm{x} \in \mathbb{R}^3$ and timestamp $t\in\mathbb{R}$ is defined as:
\begin{equation}
    G_i(\bm{x}, t) = \rho(\bm{p}_i,t) \cdot \exp{\left(-\frac{1}{2}(\bm{x}-\bm{p}_i)^{\top}\mathbf{\Sigma}_{i}^{-1}(\bm{x}-\bm{p}_i)\right)}.
\end{equation}
The overall contrast agent attenuation value $\mu(\bm{x},t)\in \mathbb{R}_{\geq 0}$ is computed as the sum of responses from all kernels:
\begin{equation}
    \mu(\bm{x},t) = \sum_{i=1}^{M}G_i(\bm{x},t).
\label{eq:contrast agent attenuation}
\end{equation}

\subsubsection{Bounded Scaling Activation}
\label{sec:bounded scaling activation}

Geometry parameters $\bm{p}_i$, $\bm{r}_i$, and $\bm{s}_i$ are activated from their optimizable raw counterparts $\tilde{\bm{p}}_i\in\mathbb{R}^3$, $\tilde{\bm{r}}_i\in\mathbb{R}^4$, and $\tilde{\bm{s}}_i\in\mathbb{R}^3$, respectively, to ensure they remain within valid ranges. Following 3DGS~\cite{3DGS}, we activate positions with $\bm{p}_i=\tilde{\bm{p}}_i$ and rotation quaternions with $\bm{r}_i=\mathrm{norm}(\tilde{\bm{r}}_i)$, where $\mathrm{norm}(\cdot)$ is the normalization operation. Regarding scale vectors, we avoid using exponential activation in~\cite{3DGS}.
Because its unbounded positive output range would cause elongated kernels, resulting in needle artifacts~\cite{3DGS}. 
In DSA imaging scenario, kernel sizes should remain small to model fine vessel details. 
Motivated by this structure prior, we adopt bounded scaling activation similar to~\cite{xu2024grm}:
\begin{equation}
    \bm{s}_i = \left(s_{max}-s_{min}\right)\sigma(\tilde{\bm{s}}_i)+s_{min},
    \label{eq:bounded_scaling_activation}
\end{equation}
where $\sigma(\cdot)$ is the sigmoid activation, and $s_{min},s_{max}\in\mathbb{R}_{+}$ are the minimum and maximum scale bounds, respectively. 
As a result, elements of $\bm{s}_i$ are constrained in the range of $(s_{min}, s_{max})$, effectively mitigating needle artifacts and enhancing reconstruction quality.

\subsubsection{Dynamic Neural Attenuation Field}
To model attenuation changes in Gaussian kernels, we introduce a compact neural network $\mathcal{D}$ dubbed dynamic neural attenuation field (DNAF). 
It takes kernel position and timestamp as input, and predicts the central attenuation: $\mathcal{D}:(\bm{p}_{i},t)\in\mathbb{R}^3\times\mathbb{R}\rightarrow \rho \in \mathbb{R}_{\geq 0}$. To enhance spatial-temporal expressiveness, we adopt the 3D hash encoding $\bm{h}_{\mathrm{3D}}$~\cite{Instant-ngp} for time-invariant features and the 4D hash encoding $\bm{h}_{\mathrm{4D}}$~\cite{4dhash} for time-variant features. A subsequent decoding multi-layer perceptron (MLP) $\phi$ then maps these features to attenuation values, activated by rectified linear unit (ReLU)~\cite{ReLU}. Mathematically, $\rho(\bm{p}_i,t)$ is formulated as:
\begin{equation}
    \rho(\bm{p}_i,t) = \phi\left(\bm{h}_{\mathrm{3D}}(\bm{p}_i)\oplus\bm{h}_{\mathrm{4D}}(\bm{p}_i,t)\right),
\end{equation}
where $\oplus$ denotes concatenation operation.

\subsubsection{Model Initialization}
Inspired by R$^2$-Gaussian~\cite{R2GS}, we initialize geometry parameters of kernels from a low-quality volume reconstructed by FDK~\cite{FDK}. 
Specifically, we sample $M$ non-empty voxels with an attenuation threshold $\delta$, and assign their locations as initial raw positions $\{\tilde{\bm{p}}_i\}_{i=1}^{M}$.
We set the activated scales $\{\bm{s}_i\}_{i=1}^{M}$ as the nearest neighbor distances of positions, and compute raw scales $\{\tilde{\bm{s}}_i\}_{i=1}^{M}$ with the inverse of~\cref{eq:bounded_scaling_activation}.
We set raw rotations as $\{\tilde{\bm{r}}_i = [1, 0, 0, 0]^{\top}\}_{i=1}^{M}$.
Network parameters in DNAF are randomly initialized.

\subsection{Model Optimization}
\label{sec:model_optimization}

\subsubsection{DSA Image Synthesis}
Consider an X-ray path $\bm{l}(a)=\bm{o}+a\bm{d}\in\mathbb{R}^3$ sampled from the ${j_k}$-th frame ray set $\mathbb{L}_{j_k}\subset\mathbb{R}^3$ for training, where $\bm{o}\in\mathbb{R}^3$, $a\in\mathbb{R}_{\geq0}$, and $\bm{d}\in\mathbb{R}^3$ are source position, length variable, and ray direction, respectively. Based on the Beer-Lambert law~\cite{BeerLaw}, the corresponding synthesized DSA pixel value $\hat{I}\in\mathbb{R}_{\geq0}$ is obtained by integrating the contrast agent attenuation $\mu$ (\cref{eq:contrast agent attenuation}) along the ray path:
\begin{equation}
\label{eq:x-ray imaging}
    \hat{I}(\bm{l},t_{j_k})=\int_{a_{n}}^{a_{f}} \mu\left(\bm{l}(a), t_{j_k}\right) \mathrm{d}a,
\end{equation}
where $[a_n, a_f]$ is the path bound. 
The complete synthesized DSA image $\hat{\mathbf{I}}(t_{j_k})\in\mathbb{R}^{H\times W}$ is then obtained by compositing the pixel values from the entire ray set $\mathbb{L}_{j_k}$: $\hat{\mathbf{I}}(t_{j_k})=\{\hat{I}(\bm{l}, t_{j_k})\}_{\bm{l}\in\mathbb{L}_{j_k}}$.
Such a synthesizing process can be efficiently achieved via X-ray rasterization~\cite{R2GS}, which splats Gaussian kernels onto the image plane and parallelly computes ray integrations.

\subsubsection{Training Objective}
We optimize our model in~\cref{sec:4d_Radiative_Gaussian_Modeling} by minimizing the L1 loss $\mathcal{L}_{1}$ and D-SSIM loss $\mathcal{L}_{ssim}$~\cite{SSIM} between the synthesized DSA images $\hat{\mathbf{I}}(t_{j_k})$ and real captured ones $\mathbf{I}_{j_k}$. To mitigate overfitting on training frames and improve temporal consistency, we follow~\cite{VPAL} to incorporate temporal perturbation into the loss function. The overall loss function is formulated as:
\begin{equation}
    \mathcal{L}=(1-\lambda_{ssim})\mathcal{L}_1\left(\hat{\mathbf{I}}(t_{j_k}+\tau), \mathbf{I}_{j_k}\right) + \lambda_{ssim}\mathcal{L}_{ssim}\left(\hat{\mathbf{I}}(t_{j_k}+\tau), \mathbf{I}_{j_k}\right), \tau\sim\mathcal{N}(0, w^2),
\end{equation}
where $w=t_{j_{k}+1}-t_{j_{k}}$ is the standard deviation of temporal Gaussian noise $\tau$.

\subsubsection{Accumulated Attenuation Pruning}
\label{Sec:accumulation_attenuation_prune}

During training, we refine the kernel distribution to better match the target vessel geometry via adaptive control mechanism including densification and pruning.
The densification strategy is the same as in 3DGS~\cite{3DGS} and R$^2$-Gaussian~\cite{R2GS}.
If a kernel's position gradient exceeds a predefined threshold, it suggests that this kernel does not accurately represent the underlying area and requires densification. 
As illustrated in \cref{fig:method}(b), small Gaussians (indicated by light blue circles) in under-reconstructed regions are densified through cloning, while large Gaussians (dark blue circles) in over-reconstructed regions are densified by splitting.

R$^2$-Gaussian~\cite{R2GS} prunes empty kernels with central attenuation values below a predefined threshold. 
However, this approach is unsuitable for our 4D kernels in DSA imaging, as their attenuation values vary over time. 
Using current timestamp attenuation as the pruning criterion may mistakenly remove vessel kernels that are not marked by the contrast agent at that moment. 
TOGS~\cite{TOGS} proposes a random pruning strategy, but it would also prune vessel kernels, resulting in degraded reconstruction.
To overcome this limitation, we introduce accumulated attenuation pruning. 
The accumulated attenuation $A_i$ of a kernel is defined as:
\begin{equation}
    A_i = \frac{\sum_{iter}\rho(\bm{p}_i,t)}{\sum_{iter}1},
\end{equation}
where $\sum_{iter}$ denotes the summation over training iterations between neighboring pruning operations. 
If the accumulated attenuation remains consistently small, i.e., $A_i <\epsilon$, it means this kernel has been negligibly marked by contrast agent during training. 
This suggests the kernel belongs to backgrounds rather than vessels, and therefore should be pruned.
In this way, we precisely prune empty kernels and retain useful ones, eventually improving reconstruction quality.

\subsection{Vessel Reconstruction}
\label{sec:vascular_reconstruction}
Given any timestamp $t$, we leverage CUDA-based voxelization in~\cite{R2GS} to efficiently query an attenuation volume $\mathbf{V}(t)=\{\mu(\bm{x},t)\}_{\bm{x}\in\mathbb{X}}$. 
Here, $\mathbb{X}\subset\mathbb{R}^3$ is the set of voxel grids defined by the target volume's resolution and spacing.
The final 3D vessel volume is obtained by averaging attenuation volumes across all timestamps $\{t_j\}_{j=1}^{T}$ in the complete set of DSA data (\cref{fig:method}(c)): $\overline{\mathbf{V}}=\frac{1}{T}\sum_{j=1}^{T}\mathbf{V}(t_j)$.

\section{Experiments}

 \subsection{Experimental Settings}

\begin{table}[t]
\centering
\fontsize{6.5}{6.5}\selectfont
\setlength{\tabcolsep}{6pt}
\renewcommand\arraystretch{1.5}
\caption{Configurations of DSA images and reconstructed volumes used in experiment.}
\label{tab:data detail}
\resizebox{\columnwidth}{!}{%
\begin{tabular}{@{}ccccc@{}}
\toprule
Cases   & Image resolution & Pixel size ($\mathrm{mm}$) & Volume resolution & Voxel size ($\mathrm{mm}$)  \\ 
\midrule
\#7     & 1240$\times$960    & 0.3144$\times$0.3173    & 512$\times$512$\times$400  & 0.4768$^3$  \\
\#10,~\#11,~\#13 & 960$\times$960   & 0.3236$\times$0.3198 & 512$\times$512$\times$506      & 0.3802$^3$ \\
\#15    & 1240$\times$960   & 0.3081$\times$0.3070  & 512$\times$512$\times$395  & 0.4663$^3$  \\
Others  & 1240$\times$960         & 0.3219$\times$0.3208 & 512$\times$512$\times$395       & 0.4881$^3$ \\ \bottomrule
\end{tabular}%
}
\end{table}

\noindent \textbf{Dataset}
In this study, we collected data from 15 real-world patient cases using the Siemens AXIOM-Artis DSA scanning system, whose source-to-object and source-to-detector distances are 750$\mathrm{mm}$ and 1200$\mathrm{mm}$, respectively.
For each case, the system captured 133 mask-fill X-ray image pairs, evenly distributed across a rotational range of 198 degrees.
Additionally, the system provided vessel volumes reconstructed by its inbuilt FDK algorithm~\cite{fdk-3ddsarecon}. Detailed configurations of DSA images and reconstructed volumes are listed in~\cref{tab:data detail}. Notably, 
although the provided volumes are not entirely accurate, we treat them as references for evaluating 3D reconstruction. We subsampled 30, 40, 50, and 60 views from the complete set as four training scenarios, and left the remaining views for evaluating 2D image synthesis.

\noindent \textbf{Implementation Details}
DNAF comprises two hash encoders $\bm{h}_{\mathrm{3D}}$ and $\bm{h}_{\mathrm{4D}}$ followed by a two-layer MLP with a width of 64.
$\bm{h}_{\mathrm{3D}}$ consists of 12 hash levels.
The hash table size is $2^{19}$ with every entry storing a trainable 2-dimensional feature vector.
The base resolution is set to 8, increasing by a factor of 1.45 per level.
$\bm{h}_{\mathrm{4D}}$ uses a similar configuration, except the base resolution is set to 2, with a growth factor of 1.4.
The rasterization and voxelization modules are borrowed from R$^2$-Gaussian~\cite{R2GS}.
We optimize our model with Adam optimizer~\cite{Adam} for 30k iterations.
Adaptive control runs from 600 to 15k iterations, adjusting every 200 iterations with a gradient threshold of 0.0001.
The threshold for accumulated attenuation pruning is $\epsilon=1e-6$.
A fast version is also provided, where the model trains for 10k iterations, and adaptive control stops at 5k iterations.
The learning rates for position, rotation, scale, and DNAF start from 0.0001, 0.001, 0.005, and 0.001, respectively, and exponentially decay to 0.1 of the initial ones.
There is also a weight decay factor of $5e-5$ for parameters in DNAF.
$s_{min}$ and $s_{max}$ are set to 0.1 and 10 times of the voxel spacing, respectively.
We initialize $M=30$k kernels with threshold $\delta=0.016$.
The loss weight is $\lambda_{ssim}=0.2$.

\noindent \textbf{Evaluation Metrics}
We evaluated both 3D vessel reconstruction and 2D DSA image synthesis.
For 3D reconstruction, we did not directly compare the reconstructed volumes with the reference volumes due to the unknown data calibration issue. Instead, we evaluated vessel surfaces, which are easier to align and compare. Specifically, we used marching cubes~\cite{marchingcube} to extract meshes from volumes, with attenuation thresholds of 0.025 for the reference volumes and 0.008 for the algorithm reconstructed volumes. The reference and reconstructed meshes were then aligned using iterative closest point (ICP)~\cite{ICP}. After that, we computed the Chamfer distance (CD) and Hausdorff distance (HD) between the aligned meshes, both measured in millimeters. For 2D image synthesis evaluation, we calculated the peak signal-to-noise ratio (PSNR) and structural similarity index measure (SSIM)~\cite{SSIM} between the synthesized and ground truth DSA images in the test set. We also recorded the running time as an efficiency metric.

\noindent \textbf{Competing Methods}
We compared 4DRGS with four methods: FDK with hann filtering~\cite{FDK,fdk-3ddsarecon}, 3DGS-based CT reconstruction method R$^2$-Gaussian~\cite{R2GS}, 3DGS-based DSA image synthesis method TOGS~\cite{TOGS}, and current SOTA NeRF-based method VPAL~\cite{VPAL}. FDK was implemented based on TIGRE-toolbox~\cite{tigre}, while others were adapted from their source codes. FDK and R$^2$-Gaussian are static methods, so we directly evaluated their output volumes. TOGS, VPAL, and our 4DRGS are dynamic methods, and we computed the average of generated volumes across timestamps as the final output, as described in~\cref{sec:vascular_reconstruction}.
TOGS does not support direct volume reconstruction.
As a workaround, we uniformly rendered 720 views in a full circle~\cite{SCOPE} using TOGS and then reconstructed the volume with FDK~\cite{fdk-3ddsarecon} for each timestamp.
All experiments were conducted on a single RTX3090 GPU.

\subsection{Experimental Results}

\subsubsection{Quantitative Evaluation}

\begin{table}[t]
\caption{Quantitative results on 3D vessel reconstruction and 2D DSA image synthesis. The best performance is shown in bold, while the second best is underlined.}
\centering
\fontsize{6.5}{6.5}\selectfont
\setlength{\tabcolsep}{5.5pt}
\renewcommand\arraystretch{1.2}
\begin{tabular}{ccccccc}
\toprule
Views               & Method         & CD ($\mathrm{mm}$) $\downarrow$          & HD ($\mathrm{mm}$) $\downarrow$          & PSNR ($\mathrm{dB}$) $\uparrow$          & SSIM $\uparrow$           & Time $\downarrow$           \\ \midrule
\multirow{6}{*}{30} & FDK            & 32.63$\pm$5.90 & 90.24$\pm$13.61 & -              & -               & \textbf{1.06s$\pm$0.20s} \\
                    & R$^2$-Gaussian & 8.34$\pm$1.93 & 34.56$\pm$8.30 & 28.61$\pm$1.22 & 0.803$\pm$0.053 & 11m55s$\pm$20s  \\
                    & TOGS           & 5.24$\pm$1.36 & 18.66$\pm$11.08 & 34.17$\pm$1.79 & 0.803$\pm$0.072 & 8m46s$\pm$22s   \\
                    & VPAL           & \underline{1.79$\pm$0.51} & 4.07$\pm$1.79 & 34.32$\pm$1.84 & 0.819$\pm$0.068 & 2h36m$\pm$52s   \\
                    & Ours (10k)& 1.88$\pm$0.36 & \underline{4.05$\pm$1.22} & \underline{34.96$\pm$1.73} & \underline{0.867$\pm$0.053} & \underline{4m31s$\pm$9s} \\
                    & Ours (30k)& \textbf{1.72$\pm$0.29} & \textbf{3.44$\pm$0.85} & \textbf{35.07$\pm$1.65} & \textbf{0.869$\pm$0.052} & 12m38s$\pm$37s  \\ \midrule 
\multirow{6}{*}{40} & FDK            & 30.61$\pm$6.12 & 87.23$\pm$15.10 & -              & -               & \textbf{1.36s$\pm$0.24s} \\
                    & R$^2$-Gaussian & 8.13$\pm$1.68 & 33.91$\pm$7.16 & 28.63$\pm$1.31 & 0.804$\pm$0.054 & 11m50s$\pm$27s  \\
                    & TOGS           & 5.04$\pm$1.35 & 18.19$\pm$9.80 & 34.43$\pm$1.83 & 0.803$\pm$0.075 & 8m49s$\pm$39s  \\
                    & VPAL           & \underline{1.69$\pm$0.40} & \underline{3.78$\pm$1.28} & 35.05$\pm$2.33 & 0.824$\pm$0.071 & 2h38m$\pm$3m54s   \\
                    & Ours (10k)& 1.81$\pm$0.30 & 3.89$\pm$1.00 & \underline{35.50$\pm$1.76} & \underline{0.871$\pm$0.053} & \underline{4m40s$\pm$10s} \\
                    & Ours (30k)& \textbf{1.67$\pm$0.27} & \textbf{3.15$\pm$0.80} & \textbf{35.69$\pm$1.77} & \textbf{0.874$\pm$0.051} & 12m33s$\pm$28s  \\ \midrule 
\multirow{6}{*}{50} & FDK            & 28.80$\pm$6.64 & 84.36$\pm$16.57 & -              & -               &\textbf{1.62s$\pm$0.33s} \\
                    & R$^2$-Gaussian & 8.07$\pm$1.77 & 34.16$\pm$8.62 & 28.70$\pm$1.30 & 0.807$\pm$0.054 & 12m2s$\pm$25s  \\
                    & TOGS           & 5.01$\pm$1.50 & 18.25$\pm$11.37 & 34.69$\pm$1.95 & 0.808$\pm$0.080 & 8m50s$\pm$38s   \\
                    & VPAL           & \textbf{1.58$\pm$0.19} & \underline{3.59$\pm$1.05} & \underline{35.91$\pm$1.83} & 0.832$\pm$0.070 & 2h44m$\pm$8m30s   \\
                    & Ours (10k)& 1.78$\pm$0.38 & 3.89$\pm$1.00 & 35.85$\pm$1.82 & \underline{0.875$\pm$0.052} & \underline{4m43s$\pm$13s} \\
                    & Ours (30k)& \underline{1.67$\pm$0.29} & \textbf{3.20$\pm$0.82} & \textbf{36.13$\pm$1.86} & \textbf{0.879$\pm$0.050} & 12m56s$\pm$29s  \\ \midrule 
\multirow{6}{*}{60} & FDK            & 27.04$\pm$7.36 & 82.33$\pm$18.08 & -              & -               &\textbf{1.89s$\pm$0.32s} \\
                    & R$^2$-Gaussian & 7.95$\pm$1.60 & 33.72$\pm$7.42 & 28.75$\pm$1.26 & 0.808$\pm$0.053 & 12m5s$\pm$18s  \\
                    & TOGS           & 4.96$\pm$1.24 & 18.16$\pm$11.10 & 34.73$\pm$1.82 & 0.812$\pm$0.067 & 9m17s$\pm$34s  \\
                    & VPAL           & \underline{1.75$\pm$0.68} & 3.94$\pm$2.17 & 35.85$\pm$2.73 & 0.830$\pm$0.074 & 2h42m$\pm$7m28s   \\
                    & Ours (10k)& 1.91$\pm$0.52 & \underline{3.70$\pm$0.90} & \underline{36.06$\pm$1.83} & \underline{0.876$\pm$0.052} & \underline{4m38s$\pm$9s} \\
                    & Ours (30k)& \textbf{1.63$\pm$0.26} & \textbf{3.27$\pm$0.83} & \textbf{36.33$\pm$1.87} & \textbf{0.880$\pm$0.051} & 12m52s$\pm$47s  \\ \bottomrule
\end{tabular}
\label{tab:results}
\end{table}

\cref{tab:results} presents a quantitative comparison of different methods.
Our method achieves SOTA 2D and 3D performance in most scenarios, with only a slightly poorer CD metric than VPAL at 50 input views. 
In terms of efficiency, our method converges in 13 minutes, offering a speedup of over 12$\times$ compared to VPAL, which requires around 160 minutes for training.
Remarkably, our fast version, trained for 10k iterations, achieves comparable performance as VPAL in 5 minutes, which is 32$\times$ faster.
Our efficiency arises from two reasons.
First, we use Gaussian kernels to model only vascular structures, avoiding unnecessary computations for empty backgrounds.
In contrast, VPAL employs MLPs to model entire scanning scenes.
Second, highly parallelized rasterization is inherently faster than VPAL's volume rendering.

\subsubsection{Qualitative Evaluation}

\begin{figure*}[t]
\centering
    \includegraphics[width=1.0\textwidth]{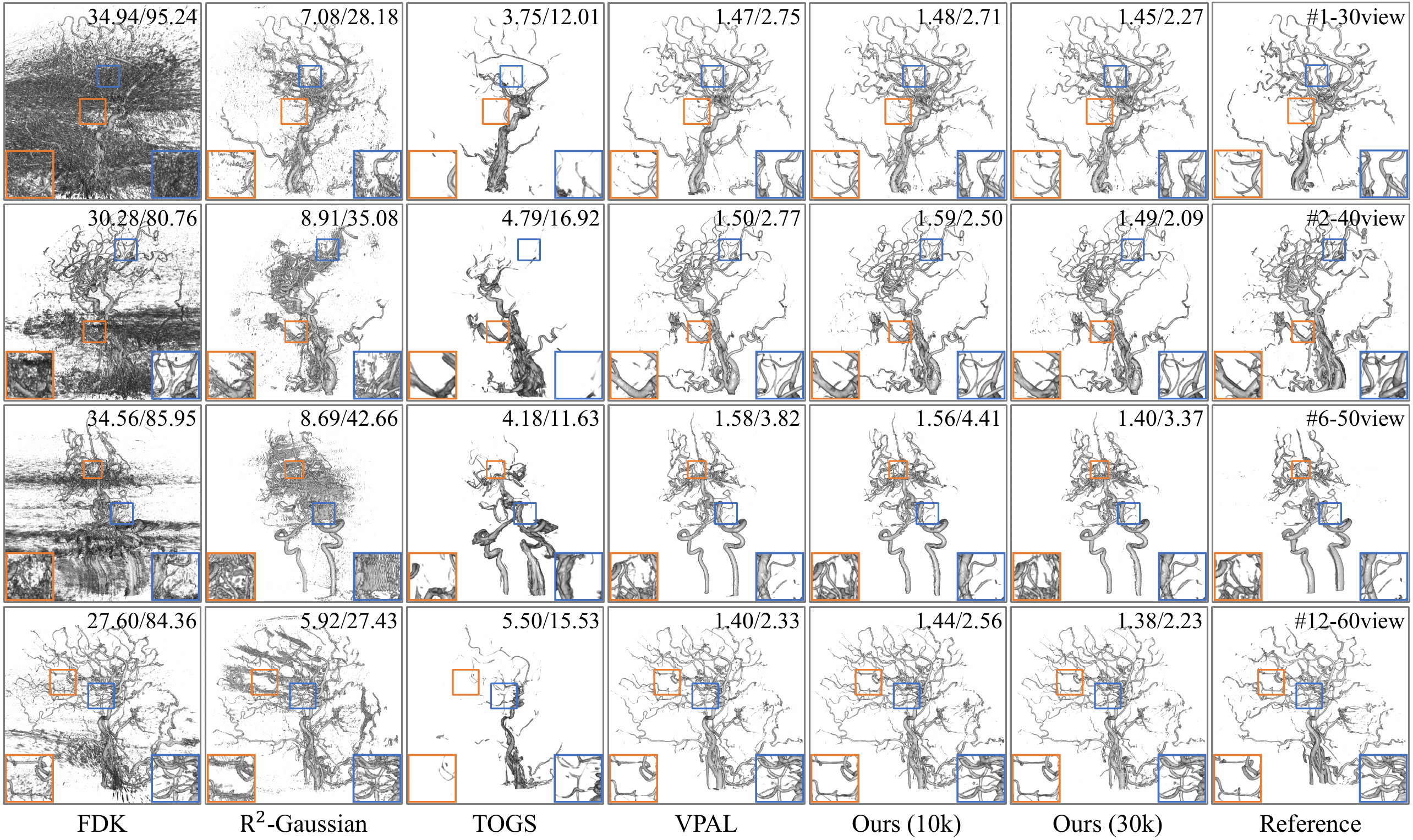}
    \caption{
    3D vessel reconstruction of different methods with CD($\mathrm{mm}$)/HD($\mathrm{mm}$) values shown at the top right of each image.
    } 
    \label{fig:3dresult}
\end{figure*}

\begin{figure*}[t]
\centering
    \includegraphics[width=1.0\textwidth]{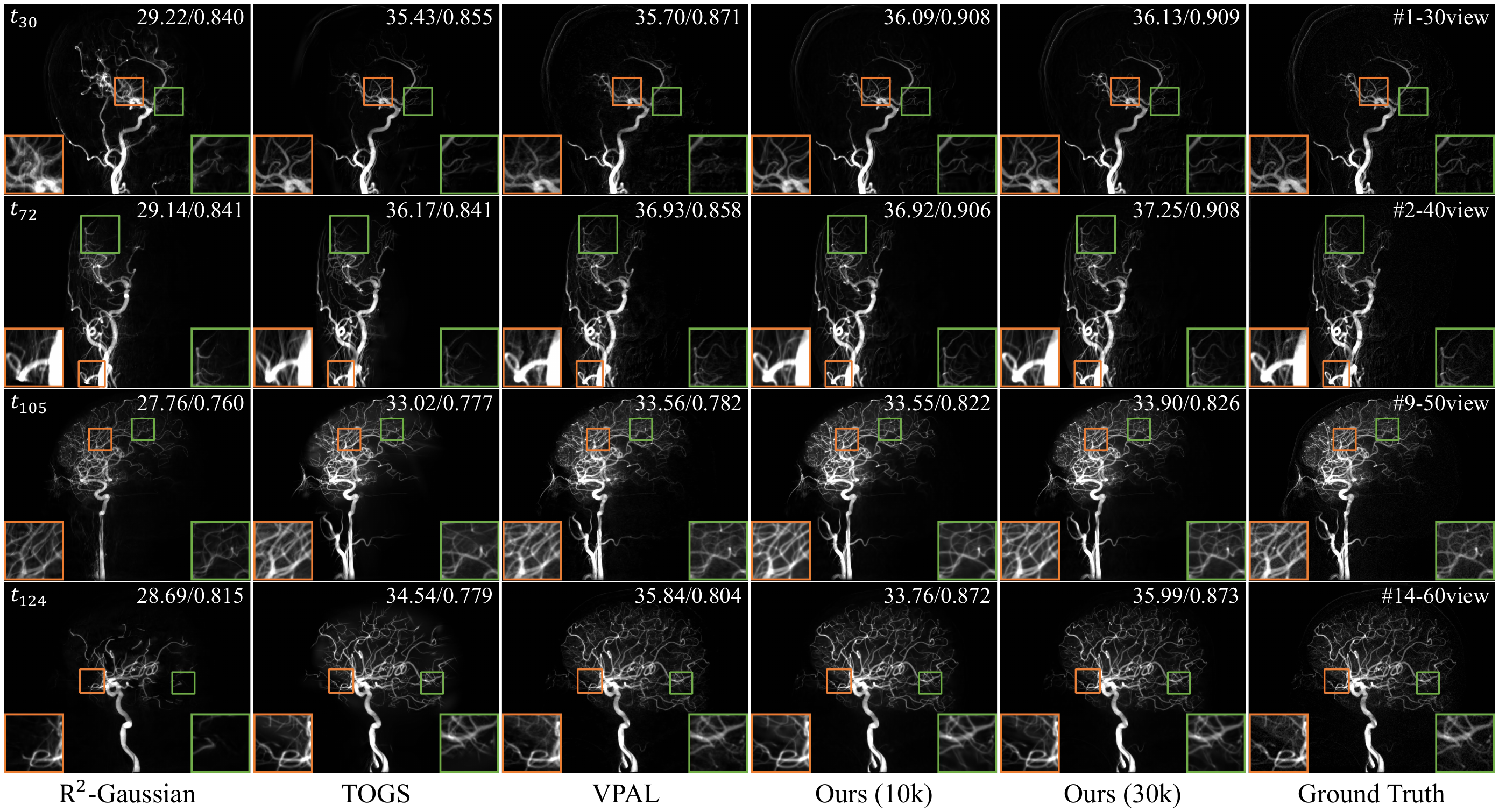}
    \caption{
    2D DSA image synthesis of different methods at test frames.
    PSNR($\mathrm{dB}$)/SSIM values averaged over the test set are shown at the top right of each image.
    } 
    \label{fig:2dresult}
\end{figure*}

A qualitative comparison of 3D vessel reconstruction is shown in \cref{fig:3dresult}.
Two static methods, FDK and R$^2$-Gaussian, exhibit severe streaky artifacts and noises, highlighting the necessity of dynamic flow modeling. TOGS generates blurry reconstructions because it trivially adapts RGB rasterization from original 3DGS to DSA imaging without considering X-ray formation principles. While VPAL recovers complete vessel structures, it still fails to capture some fine details. In contrast, our method effectively preserves intricate vessel details, demonstrating the superiority of our kernel-based representation compared to the pure MLP-based representation used in VPAL.

\cref{fig:2dresult} presents a qualitative comparison of 2D DSA image synthesis on test frames. 
R$^2$-Gaussian loses significant vessel structures due to its inability to model dynamic DSA sequences. 
TOGS performs better but vessel details remain under-reconstructed.
Because its limited temporal modeling hinders accurate recovery of contrast agent dynamics.
VPAL renders complete structures but lacks fine details, and suffers from noise and blurriness.
In contrast, our method captures fine details while minimizing artifacts and noise, showcasing its effectiveness in vessel structure representation.

\subsection{Ablation Study}

In this section, we conduct ablation studies to validate our two innovations: accumulated attenuation pruning and bounded scaling activation.
Accumulated pruning is compared to random pruning in TOGS and threshold pruning in R$^2$-Gaussian. 
We follow the original papers to set the random pruning proportion to 8\% and the threshold pruning attenuation to $1e-6$.
Bounded activation is compared with exponential activation in vanilla 3DGS.
All ablation experiments are conducted with 30 input views for 30k training iterations.

\begin{table}[t]
\caption{Quantitative results of ablation study. \textbf{RP} stands for random pruning, \textbf{TP} stands for threshold pruning, \textbf{Exp} stands for exponential scaling activation, and \textbf{Gau.} stands for number of Gaussians.
 }
\centering
\fontsize{7.5}{7.5}\selectfont
\setlength{\tabcolsep}{6pt}
\renewcommand\arraystretch{1.5}
\begin{tabular}{ccccccc}
\toprule
       & CD ($\mathrm{mm}$) $\downarrow$ & HD ($\mathrm{mm}$) $\downarrow$ & PSNR ($\mathrm{dB}$) $\uparrow$ & SSIM $\uparrow$ & Time $\downarrow$ & Gau. \\ 
       \midrule
RP     & 6.79$\pm$1.89  & 28.84$\pm$11.59 & 34.64$\pm$1.60   & 0.866$\pm$0.052   & 12m27s$\pm$40s   & 93k$\pm$27k  \\
TP     & 5.46$\pm$1.85  & 28.41$\pm$13.42 & 34.76$\pm$1.61   & 0.866$\pm$0.052   & 12m44s$\pm$36s   & 76k$\pm$29k  \\ \midrule 
Exp    & 2.06$\pm$0.41  & 4.04$\pm$0.94 & 35.05$\pm$1.68   & 0.870$\pm$0.052   & 13m5s$\pm$41s   & 61k$\pm$19k  \\ \midrule
Ours   & 1.72$\pm$0.29  & 3.44$\pm$0.85 & 35.07$\pm$1.65   & 0.869$\pm$0.052   & 12m38s$\pm$37s   & 61k$\pm$19k  \\ \bottomrule 
\end{tabular}
\label{tab:ablation}
\end{table}

\begin{figure*}[t]
\centering
    \includegraphics[width=1.0\textwidth]{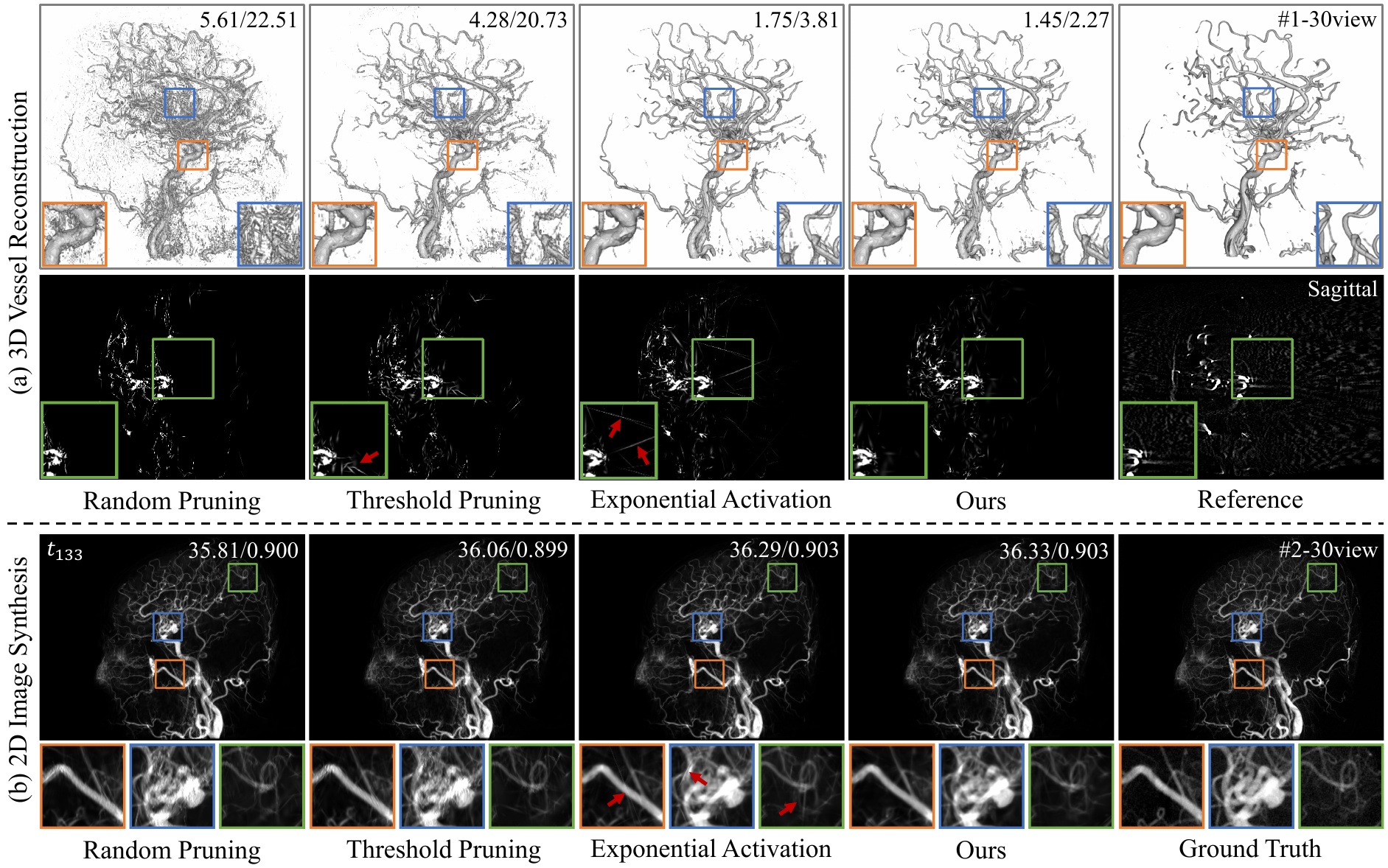}
    \caption{
    Qualitative results of ablation study. 
    (a) 3D vessel reconstruction. 
    Top row: 3D visualization with CD($\mathrm{mm}$)/HD($\mathrm{mm}$) values shown at the top right of each image.
    Bottom row: sagittal slice of reconstructed volume.
    (b) 2D DSA image synthesis at test frame.
    PSNR($\mathrm{dB}$)/SSIM values averaged over the test set are shown at the top right of each image.
    } 
    \label{fig:result_ab}
\end{figure*}

Quantitative results in \cref{tab:ablation} show that random pruning and threshold pruning lead to significant degradation of both 3D and 2D quality.
Additionally, kernel numbers tend to increase under these strategies.
Exponential activation causes a minor reduction in 3D quality, and it slightly extends the training time.
In \cref{fig:result_ab}(a), we provide 3D vessel reconstruction results at the top row and the corresponding sagittal slices at the bottom row.
Random pruning and threshold pruning introduce considerable noisy artifacts because they incorrectly prune kernels that belong to vessels.
In contrast, our accumulated pruning method precisely removes non-vessel Gaussians while preserving vessel-related ones, enabling high-quality reconstruction with limited noise.
Exponential activation produces needle artifacts as highlighted by the red arrows. 
These artifacts stem from elongated Gaussian kernels caused by the unbounded positive range of exponential activation.
In contrast, our bounded scaling activation addresses this issue by constraining the Gaussian kernel size within a suitable range.
In \cref{fig:result_ab}(b), we provide 2D DSA image synthesis results at test frames.
Consistent with 3D results, random pruning and threshold pruning exhibit noticeable quality degradation with noise, while exponential activation introduces needle artifacts.
Overall, our full model achieves the best results both quantitatively and qualitatively, verifying the effectiveness of accumulated attenuation pruning and bounded scaling activation.

\section{Discussion and Conclusion}

\noindent \textbf{Discussion}
While our method achieves SOTA results, it has some limitations. 
For instance, we assume no patient movement during scanning, though minor motion may occur in practice. 
Additionally, we have not considered calibration errors in scanner geometry and imaging noises in DSA scanning process.
Moreover, our dataset is a small private set of multi-view cerebral DSA scans.
However, single- or bi-plane DSA is more common especially for other body parts like coronary arteries.
Our method would fail on such data due to limited views or unmodeled cardiac motion.
We leave these aspects for future work.

\noindent \textbf{Conclusion}
In this work, we present 4DRGS, the first Gaussian splatting-based framework for sparse-view DSA reconstruction. 4DRGS represents vessels using 4D radiative Gaussian kernels, which effectively model static vessel structures and time-varying attenuation changes. We train these kernels with image losses and extract the target vessel volume from them. 
Two innovations are proposed to enhance reconstruction quality, including accumulated attenuation pruning and bounded scaling activation.
Extensive experiments demonstrate the superiority of our method in both 3D vessel reconstruction and 2D DSA image synthesis. Remarkably, our method achieves impressive results within minutes, highlighting its potential for clinical applications.

\begin{credits}
\subsubsection{\ackname} 
This work was supported in part by NSFC grants (No. 6230012077), and HPC Platform of ShanghaiTech University. Shanghai Municipal Central Guided Local Science and Technology Development Fund Project no: YDZX20233100001001.
\end{credits}

%
%
\bibliographystyle{splncs04}
\bibliography{reference}

\end{document}